\documentclass[journal,draftcls,onecolumn,12pt,twoside]{IEEEtranTCOM}
\normalsize
\usepackage{epsfig,cite,color}
\usepackage{amsmath,amssymb,setspace}
\usepackage{multirow,array}
\newtheorem{proposition}{Proposition}
\newtheorem{lemma}{Lemma}
\newtheorem{remark}{Remark}

\usepackage{epsfig}\usepackage{cite}\usepackage{multirow}\usepackage{array}\usepackage{subfigure}

\begin{document}

\title{Lossy Computing of Correlated Sources with Fractional Sampling}
\author{Xi~Liu,~\IEEEmembership{Student Member,~IEEE,}
        Osvaldo~Simeone,~\IEEEmembership{Member,~IEEE,}
        and~Elza~Erkip,~\IEEEmembership{Fellow,~IEEE}
\thanks{X. Liu is with Broadcom, Matawan, NJ 07747, USA (Email: xi.liu1984@gmail.com).}
\thanks{O. Simeone is with the ECE Department, New Jersey Institute of Technology, Newark, NJ 07102, USA (Email: osvaldo.simeone@njit.edu).}
\thanks{E. Erkip is with the ECE Department, Polytechnic Institute of New York University, Brooklyn, NY 11201, USA (Email: elza@poly.edu).}
}

\maketitle
\vspace{-1cm}
\begin{abstract}
This paper considers the problem of lossy compression for the computation
of a function of two correlated sources, both of which are observed
at the encoder. Due to presence of observation costs, the encoder
is allowed to observe only subsets of the samples from both sources,
with a fraction of such sample pairs possibly overlapping. The rate-distortion
function is characterized for memoryless sources, and then specialized to Gaussian and binary sources
for selected functions and with quadratic
and Hamming distortion metrics, respectively. The optimal measurement overlap fraction is shown
to depend on the function to be computed by the decoder, on the source statistics, including the
correlation, and on the link rate. Special cases are discussed in which
the optimal overlap fraction is the maximum or minimum possible value
given the sampling budget, illustrating non-trivial performance trade-offs
in the design of the sampling strategy. Finally, the analysis is extended to the multi-hop set-up with jointly Gaussian sources, where each encoder can observe only one of the sources.
\end{abstract}
\vspace{-0.5cm}
\section{Introduction}
A battery-limited wireless sensor node consumes energy in both its channel coding and its source coding components. The energy expenditure of the channel coding component is due to the power amplifier and to processing steps related to communication; instead, the source coding component consumes energy in the process of digitizing the information sources of
interest through a cascade of acquisition, sampling, quantization and compression. It is also
known that the overall energy spent for compression is generally comparable to that used for
communication and that a joint design of compression and transmission is critical to improve the
energy efficiency \cite{processingcost:Barr_Asanovic06} \cite{processingcost:Sadler_Martonosi06}. We refer to the energy associated with the source coding component, i.e.,  measurements and compression
of sources, as ``sensing energy''.

A reasonable, and analytically tractable model, for the sensing energy is obtained by assuming that the sensing cost is proportional to the number of source samples measured and compressed.\footnote{Compression schemes with close to linear complexity include Lempel-Ziv strategies and related approaches \cite{processingcost:Cover91} \cite{processingcost:Kontoyiannis_IT09}.} In our previous work \cite{processingcost:Liu_Simeone_Erkip_tcom}, in the presence of constant per-sample sensing energy, we have investigated the problem of minimizing the distortion of reconstruction of independent Gaussian sources measured by a single integrated sensor under energy
constraints on the channel and source coding components. Reference \cite{processingcost:Liu_Simeone_Erkip_tcom} reveals that, similar to the channel coding counterpart set-up in \cite{processingcost:Massaad_Medard_Zheng04}, it is generally optimal to measure and process a fraction of the source samples. We observe that this principle also underlies the compressive sensing framework \cite{processingcost:Donoho_IT06}. In this work, instead, we consider a set-up with  functional reconstruction requirements on correlated measured sources, as explained next.

\begin{figure}[ht]
 \centering \includegraphics[width=2.4in]{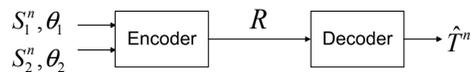} \caption{The encoder measures correlated sources $S_{1}$ and $S_{2}$
for a fraction of time $\theta_{1}$ and $\theta_{2}$, respectively,
and the decoder estimates a function $T^{n}=f^{n}(S_{1}^{n},S_{2}^{n})$.}
\label{fig:model}
\end{figure}

\begin{figure}[ht]
\centering
\includegraphics[scale=0.42]{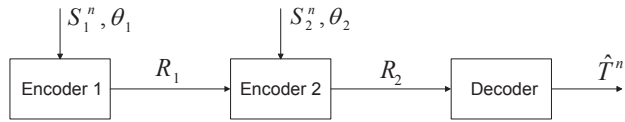}
\caption{The multi-hop setup studied in Section VII: Encoder 1 and Encoder 2 measure correlated sources $S_1$ and $S_2$ for a fraction of time $\theta_{1}$ and $\theta_{2}$, respectively, and the decoder estimates a function $T^{n}=f^{n}(S_{1}^{n},S_{2}^{n})$.}\label{fig:Cascademodel}
\end{figure}

Consider an encoder endowed with an integrated sensor that is able to measure
two correlated discrete memoryless source sequences $S_{1}^{n}=(S_{1,1},...,S_{1,n})$
and $S_{2}^{n}=(S_{2,1},...,S_{2,n})$ through two different sensor interfaces, as shown in Fig. \ref{fig:model}. Following \cite{processingcost:Liu_Simeone_Erkip_tcom}, we assume that measuring
each sample of source $S_k$, $k=1,2$, entails a constant sensing energy cost per source sample. For simplicity, instead of having a total sensing energy budget for all the sources as in \cite{processingcost:Liu_Simeone_Erkip_tcom}, we assume that the integrated sensor has a separate sensing energy budget (and thus a separate sampling budget) for either source. That is, the encoder can only measure $n\theta_{k}$ samples from source $S_{k}$, $k=1,2$, with $0\leq\theta_{k}\leq1$. The encoder compresses the measured samples to $nR$ bits, where $R$
is the communication rate in bits per source sample. Based on the
received bits, the decoder reconstructs a lossy version of a target
function $T^{n}=f^{n}(S_{1}^{n},S_{2}^{n})$ of source sequences $S_{1}^{n}$
and $S_{2}^{n}$, which is calculated symbol-by-symbol as $T_{i}=f(S_{1,i},S_{2,i})$,
$i=1,...,n$. We refer to the above problem as \textit{lossy computing
with fractional sampling}. In Section \ref{sec:multi-hop}, we will also consider the problem of \textit{multi-hop lossy computing
with fractional sampling}, which, as shown in Fig. \ref{fig:Cascademodel}, differs from the integrated sensor (point-to-point) problem in that sources $S_1$ and $S_2$ are measured by two distributed sensors connected by a finite-capacity link.

\vspace{-0.5cm}
A key aspect of the problem of \textit{\emph{lossy computing with
fractional sampling}} is that the encoder is allowed to choose \emph{which}
samples to measure given the sampling budget ($\theta_{1},\theta_{2}$).
To fix the ideas, assume that we have $(\theta_{1},\theta_{2})=(0.5,0.5)$,
so that only half of the samples can be observed from both sources.
As two extreme strategies, the encoder can either measure the same
samples from both sources, say $S_{1,i},S_{2,i}$ for $i=1,...,n/2$,
or it can measure the first source $S_{1}$ for the first $n/2$
samples, namely $S_{1,i}$ for $i=1,...,n/2$, and the second source
$S_{2}$ for the remaining $n/2$ samples, namely $S_{2,i}$ for
$i=n/2+1,...,n$. With the first sampling strategy, the encoder is
able to directly calculate the desired function $T_{i}=f(S_{1,i},S_{2,i})$
for $i=1,...,n/2$, while having no information (beside the prior
distribution) about $T_{i}$ for the remaining samples. With the second
strategy, instead, the encoder collects partial information about
 $T$ at all times in the form of samples from source
$S_{1}$ or source $S_{2}$.

Relating the discussed fractional sampling model with prior literature, we observe that, with full sampling of both sources, i.e., ($\theta_{1}=1,\theta_{2}=1$),
the encoder can directly calculate the function $T^{n}=f^{n}(S_{1}^{n},S_{2}^{n})$
and the problem at hand reduces to the standard rate-distortion set-up
(see, e.g., \cite{processingcost:Cover91}). Instead, if the encoder
can only measure one of the two sources, i.e., ($\theta_{1}=1,\theta_{2}=0$)
or ($\theta_{1}=0,\theta_{2}=1$), the problem at hand becomes a special
case of the indirect source coding set-up introduced in \cite{processingcost:Witsenhausen_IT80}. The model of fractional sampling is also related to that of compression with actions of \cite{processingcost:Zhao_Chia_Weissman_arxiv12}, in which the decoder obtains side information by taking cost-constrained actions based on the message received from the encoder. Finally, various recent information-theoretic results on the functional reconstruction problem without sampling constraints can be found in \cite{processingcost:Feizi_Medard_Allerton09} (see also references therein).

The main contributions and the organization of the paper are as follows. We formulate the problem of lossy computing with fractional
sampling of correlated sources for the set-up in Fig. \ref{fig:model} in Section II. After providing general
expressions for the distortion-rate and the rate-distortion functions in Section III, we specialize them to Gaussian sources and weighted sum function $T=w_{1}S_{1}+w_{2}S_{2}$ in Section IV and binary sources with arbitrary functions
$T=f(S_{1},S_{2})$ in Section V. As a result, various conclusions
are drawn regarding conditions under which the optimal sampling strategy
prescribes the maximum or the minimum possible overlap between the
samples measured from the two sources. In Section VI, we extend the analysis to the multi-hop set-up of Fig. \ref{fig:Cascademodel}, in which sources $S_1$ and $S_2$ are measured by different encoders connected by a finite-capacity link.

\section{System Model}

\begin{figure}[ht]
 \centering \includegraphics[width=2.5in]{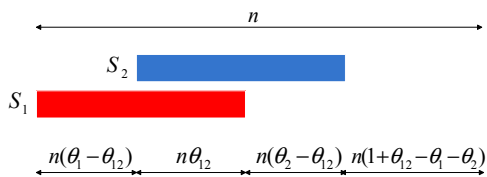} \caption{Sampling profile $(\theta_{1},\theta_{2},\theta_{12})$ at the encoder:
a fraction, $\theta_{1}-\theta_{12}$, of samples is measured only
from source $S_{1}$; a fraction, $\theta_{12}$, of samples is
measured from both sources; a fraction, $\theta_{2}-\theta_{12}$,
of samples is measured only from source $S_{2}$; and the remaining
fraction, $1+\theta_{12}-\theta_{1}-\theta_{2}$, of samples is not
measured for either source ($0\leq\theta_{1},\theta_{2}\leq1$, and
$\theta_{12}$ as in (\ref{eqn:cnst_theta12})).}
\label{fig:fraction}
\end{figure}

In this section, we formally introduce the system model of interest for the point-to-point set-up of Fig. \ref{fig:model}. The multi-hop model will be introduced in Section VII. As shown in Fig. \ref{fig:model}, the encoder has access to two discrete
memoryless source sequences $S_{1}^{n}=(S_{1,1},...,S_{1,n})$ and $S_{2}^{n}=(S_{2,1},...,S_{2,n})$
respectively, which consist of $n$ independent and identically distributed
(i.i.d.) samples $(S_{1,i},S_{2,i})$ with $S_{1,i}\in\mathcal{S}_{1}$
and $S_{2,i}\in\mathcal{S}_{2}$, $i=1,...,n$, where $\mathcal{S}_{1}$
and $\mathcal{S}_{2}$ are the alphabet sets for $S_{1}$ and $S_{2}$
respectively. All alphabets are assumed to be finite unless otherwise
stated. Due to presence of observation costs, we assume the encoder
can only sample a fraction $\theta_{k}$ of the samples for source
$S_{k}$, with $0\leq\theta_{k}\leq1$ for $k=1,2$, where the samples are
determined prior to the observation of $(S_1^n, S_2^n)$. Given the
i.i.d. nature of the sources, without loss of generality, we assume
that the encoder measures the first $\theta_{1}$ fraction of samples
of source $S_{1}$ and measures the $\theta_{2}$ fraction of
samples of $S_{2}$ starting from sample $n(\theta_{1}-\theta_{12})+1$%
\footnote{Throughout the paper, quantities such as $n\theta_{1}$, $n\theta_{2}$
and $n(\theta_{1}+\theta_{2}-\theta_{12})$ are implicitly assumed
to be rounded to the largest smaller integer.%
}, as shown in Fig. \ref{fig:fraction}. The samples measured at the
encoder from the two sources thus overlap for a fraction $\theta_{12}$,
with $\theta_{12}$ satisfying
\begin{equation}\vspace{-0.15cm}
\theta_{12,min}\leq\theta_{12}\leq\theta_{12,max},\label{eqn:cnst_theta12}
\vspace{-0.15cm}\end{equation}
with $\theta_{12,min}=(\theta_{1}+\theta_{2}-1)^{+}$ and $\theta_{12,max}=\min(\theta_{1},\theta_{2})$,
where $(\cdot)^{+}$ denotes $\max(\cdot,0)$. We refer to the triple
$(\theta_{1},\theta_{2},\theta_{12})$ as a \textit{sampling profile},
and to $(\theta_{1},\theta_{2})$ as the \emph{sampling budget}.
\vspace{-0.75cm}
The decoder wishes to estimate a function $T^{n}=f^{n}(S_{1}^{n},S_{2}^{n})$,
where $T_{i}=f(S_{1,i},S_{2,i})$ for $i=1,...,n$. We let $d:\mathcal{\textrm{ }T}\times\hat{\mathcal{T}}\rightarrow[0,+\infty)$
be a distortion measure, where $\mathcal{T}$ and $\hat{\mathcal{T}}$
are the alphabet sets of the variables $T$ and $\hat{T}$ respectively.
We assume, without loss of generality, that for each $t\in\mathcal{T}$
there exists a $\hat{t}\in\mathcal{T}$ such that $d(t,\hat{t})=0$.
The link between the encoder and the decoder can support a rate of
$R$ bits/sample. Formal definitions follow.
\vspace{-0.25cm}

\textit{Definition 1}: A $(n,R,D,\theta_{1},\theta_{2},\theta_{12})$
code for the problem of lossy computing of two memoryless sources
with fractional sampling consists of an encoder $h:\mathcal{\textrm{ }S}_{1}^{n\theta_{1}}\times\mathcal{S}_{2}^{n\theta_{2}}\rightarrow\{1,...,2^{nR}\}$,
which maps the measured $\theta_{1}$-fraction of source $S_{1}$,
i.e., $(S_{1,1},...,S_{1,n\theta_{1}})$, and the measured $\theta_{2}$-fraction
of source $S_{2}$, i.e., $(S_{2,n(\theta_{1}-\theta_{12})+1},...,$
$S_{2,n(\theta_{1}+\theta_{2}-\theta_{12})})$, into a message of
rate $R$ bits per source sample (where the normalization is with
respect to the overall number of samples $n$); and a decoder $g:\textrm{ }\{1,...,2^{nR}\}\rightarrow\hat{\mathcal{T}}^{n},$
which maps the message from the encoder into an estimate $\hat{T}^{n}$,
such that the \textit{average} distortion constraint $D$ is satisfied, i.e.,
\begin{equation}\vspace{-0.15cm}
\frac{1}{n}E\left[\sum_{i=1}^{n}d(T_{i},\hat{T}_{i})\right]\leq D.\label{eqn:AvgDist0}
\vspace{-0.15cm}\end{equation}

\textit{Definition 2}: Given any sampling profile $(\theta_{1},\theta_{2},\theta_{12})$,
a tuple $(R,D,\theta_{1},\theta_{2},\theta_{12})$ is said to be achievable,
if for any $\epsilon>0$, and sufficiently large $n$, there exists
a $(n,R,D+\epsilon,\theta_{1},\theta_{2},\theta_{12})$ code. The
\textit{distortion-rate function for sampling profile} $(\theta_{1},\theta_{2},\theta_{12})$, $D(R,\theta_{1},\theta_{2},\theta_{12})$,
is defined as $D(R,\theta_{1},\theta_{2},\theta_{12})=\inf\{D$: $(R,D,\theta_{1},\theta_{2},\theta_{12})$ is achievable$\}$, and the minimum achievable distortion for the same sampling profile is defined as $D_{min}(\theta_1,\theta_2,\theta_{12}) = \lim_{R\rightarrow \infty} D(R,\theta_1,\theta_2,\theta_{12})$.

\textit{Definition 3}: The \textit{distortion-rate function with sampling budget}
$(\theta_{1},\theta_{2})$, $D(R,\theta_{1},\theta_{2})$, is defined
as $D(R,\theta_{1},\theta_{2})=\min_{\theta_{12}}D(R,\theta_{1},\theta_{2},\theta_{12})$,
where the minimum is taken over all $\theta_{12}$ satisfying (\ref{eqn:cnst_theta12}). Moreover, the minimum achievable distortion for the same sampling budget is defined as $D_{min}(\theta_{1},\theta_{2})=\min_{\theta_{12}}D_{min}(\theta_{1},\theta_{2},\theta_{12})$.

Similar definitions are used for the rate-distortion function. Specifically,
the \textit{rate-distortion function for sampling profile} $(\theta_{1},\theta_{2},\theta_{12})$, $R(D,\theta_{1},\theta_{2},\theta_{12})$, is defined as $R(D,\theta_{1},\theta_{2},\theta_{12})=\inf\{R$:
$(R,D,\theta_{1},\theta_{2},\theta_{12})$ is achievable$\}$,
and the \textit{rate-distortion function with sampling budget} $(\theta_{1},\theta_{2})$, $R(D,\theta_{1},\theta_{2})$, is defined
as $R(D,\theta_{1},\theta_{2})=\min_{\theta_{12}}R(D,\theta_{1},\theta_{2},\theta_{12})$
where the minimum is taken over all $\theta_{12}$ satisfying (\ref{eqn:cnst_theta12}).

{\remark In most of the paper, we consider the average distortion criterion (\ref{eqn:AvgDist0}), following standard considerations, the results presented herein hold also under the definition of distortion level $D$ whereby the probability that the distortion level $D$ is exceeded by an arbitrarily small amount $\epsilon$ vanishes as the block length $n$ grows large (i.e., $Pr[(1/n) \sum_{i=1}^{n}d(T_{i},\hat{T}_{i})\geq D+\epsilon]\rightarrow 0$ as $n\rightarrow \infty$) \cite{processingcost:Marton_IT74}. Moreover, the worst-case average per-sample criterion $\max_{i\in\{1,...,n\}}E[d(T_i,\hat{T}_i)]$ is briefly considered in Appendix \ref{app:worst}.}

\vspace{-0.25cm}
\section{Rate-Distortion Trade-Off with Fractional Sampling}

\vspace{-0.25cm}In this section, we characterize the distortion-rate functions $D(R,\theta_{1},\theta_{2},\theta_{12})$
and $D(R,\theta_{1},\theta_{2})$ defined above as well as their rate-distortion
counterparts. To elaborate, we first provide some standard definitions.

\textit{Definition 4:} The standard distortion-rate function for source $T$, $D_{12}(R)$, is defined as $D_{12}(R)=\min_{p(\hat{t}|t):\textrm{ }I(T;\hat{T})\leq R}E[d(T,\hat{T})]$ \cite{processingcost:Cover91}. Moreover, the indirect distortion-rate function for compression
of source $T$ when only source $S_{k}$ is observed at the encoder, $D_{k}(R)$, is $D_{k}(R)=\min_{p(\hat{t}|s_k):\textrm{ }I(S_{k};\hat{T})\leq R}$ $E[d(T,\hat{T})]$. The distortion $D_{k,min}$ is defined as $D_{k,min} = \lim_{R\rightarrow \infty} D_k(R)= \min_{g_k(\cdot)}E(d(T,$ $g_k(S_k)))$, for $k = 1,2$, with  $g_k:\mathcal{\textrm{ }S}_{k}\rightarrow\hat{\mathcal{T}}$. Finally, the distortion $D_{max}$ is $D_{max}=\min_{\hat{t}\in\hat{\mathcal{T}}}E[d(T,\hat{t})]$.

We similarly define the corresponding rate-distortion functions $R_{12}(D)$ and $R_{k}(D)$, $k = 1,2$.

\vspace{-0.25cm}{\lemma \label{lem:lem1} For any given sampling profile $(\theta_{1},\theta_{2},\theta_{12})$
and link rate $R$, the distortion-rate function for computing $T$
is given by%
\footnote{For the distortion function $D_1(x)$ for $x\geq0$, we define $0\cdot D_1(x/0)$
= 0, for $x\geq0$, if $\lim_{x\rightarrow0}x\cdot D_1(1/x)=0$, and similarly for the distortion functions $D_{12}(x)$ and $D_2(x)$.}
\vspace{-0.25cm}\begin{align}\vspace{-0.15cm}
D(R, \theta_{1},\theta_{2},\theta_{12})=&\min_{R_{1},R_{2},R_{12}\geq0}(\theta_{1}-\theta_{12})D_{1}\left(\frac{R_{1}}{\theta_{1}-\theta_{12}}\right) +\theta_{12}D_{12}\left(\frac{R_{12}}{\theta_{12}}\right)\nonumber\\
&+(\theta_{2}-\theta_{12})D_{2}\left(\frac{R_{2}}{\theta_{2}-\theta_{12}}\right) +(1+\theta_{12}-\theta_{1}-\theta_{2})D_{max},\label{eqn:D_lemma1}
\vspace{-0.15cm}\end{align}
\vspace{-0.25cm}where the minimization is taken under the constraint $R_{1}+R_{2}+R_{12}\leq R$, and the minimum achievable distortion is given by\vspace{-0.25cm}
\begin{equation}\vspace{-0.25cm}
D_{min}(\theta_1,\theta_2,\theta_{12}) = (\theta_{1}-\theta_{12})D_{1,min} + (\theta_{2}-\theta_{12})D_{2,min}+ (1+\theta_{12}-\theta_{1}-\theta_{2})D_{max}.\label{eqn:Dmin_theta12}
\vspace{-0.25cm}\end{equation} \vspace{-0.25cm}Moreover, for any given sampling profile $(\theta_{1},\theta_{2},\theta_{12})$
and distortion level $D\geq D_{min}(\theta_1,\theta_2,\theta_{12})$, the rate-distortion function for computing
$T$ is given by \vspace{-0.25cm}\begin{align}\vspace{-0.25cm}
R(D, \theta_{1},\theta_{2},\theta_{12})=&\min_{D_{1},D_{2},D_{12}}(\theta_{1}-\theta_{12})R_{1}\left(\frac{D_{1}}{\theta_{1}-\theta_{12}}\right)\nonumber\\
 & +\theta_{12}R_{12}\left(\frac{D_{12}}{\theta_{12}}\right)+(\theta_{2}-\theta_{12})R_{2}\left(\frac{D_{2}}{\theta_{2}-\theta_{12}}\right),\label{eqn:RD_lemma1}
\end{align}
where the minimization is taken over all choices of $D_{1}$, $D_{2}$
and $D_{12}$ satisfying $D_{12}\geq0$, $D_{k}\geq(\theta_{k}-\theta_{12})D_{k,min}$, $k = 1,2$, and
\begin{equation} D_{1}+ D_{2}+D_{12}+(1+\theta_{12}-\theta_{1}-\theta_{2})D_{max}\leq D.\label{eqn:dist_sum}\end{equation}}
\vspace{-0.5cm}
\begin{proof}
The rate-distortion function (\ref{eqn:RD_lemma1}), and the corresponding distortion-rate function (\ref{eqn:D_lemma1}) can be obtained by noting that the rate-distortion problem with fractional sampling at hand can in fact be viewed as a special case of the conditional rate-distortion problem \cite{processingcost:Gray72}. To this end, let $Q\in \{1,2,3,4\}$ be a time-sharing random variable independent of $S_1$ and $S_2$ and distributed as: $Pr(Q=1) = \theta_1 - \theta_{12}$, $Pr(Q=2) = \theta_2 - \theta_{12}$, $Pr(Q=3) = \theta_{12}$, $Pr(Q=4) = 1+\theta_{12}-\theta_1-\theta_2$. Also, let $X= S_1$ if $Q = 1$, $X=S_2$ if $Q = 2$, $X = (S_1,S_2)$ if $Q = 3$, $X = \text{constant}$ if $Q = 4$. For any given sampling profile $(\theta_1,\theta_2,\theta_{12})$, the rate-distortion problem at hand reduces to a standard Wyner-Ziv problem  with $(X,Q)$ as the source available at the encoder and $Q$ as side information available both at the encoder and the decoder. Hence, the rate-distortion function in (5) is given as $R(D) = \min_{p(\hat{t}|q,x)} I(X;\hat{T}|Q)$, where the minimum is taken over the set of all conditional distributions $p(\hat{t}|q,x)$ for which the joint distribution $p(q,x,\hat{t}) = p(q)p(x|q)p(\hat{t}|q,x)$ satisfies the expected distortion constraint $E [d(T,\hat{T})]\leq D$.  This expression can be easily evaluated to (\ref{eqn:RD_lemma1}).

Note that the number of samples for each fraction in Fig. \ref{fig:fraction} grows to infinity for $n\rightarrow \infty$ as long as its corresponding fraction (i.e., $\theta_1-\theta_{12}$, $\theta_{12}$, $\theta_2 - \theta_{12}$ or $1+\theta_{12}-\theta_{1}-\theta_{2}$) is non-zero. Therefore, these fractions can be considered separately without loss of optimality.
\end{proof}
Note that in Lemma \ref{lem:lem1}, rate $R_{k}$ is assigned for the description
of the $(\theta_{k}-\theta_{12})$-fraction of samples in which only
source $S_{k}$ is measured, $k=1,2$, while rate $R_{12}$ is
assigned for the description of the $\theta_{12}$-fraction of samples
in which both sources are measured (recall Fig. \ref{fig:fraction}).
Distortions $D_{1}$, $D_{2}$ and $D_{12}$ are the
weighted average per-sample distortions in the reconstruction of $T$ at
the decoder for the corresponding fractions of samples.
\vspace{-0.25cm}{\lemma \label{lem:lem_Dmin} For any given sampling budget $(\theta_{1},\theta_{2})$, the minimum achievable distortion for computing $T = f(S_{1},S_{2})$ is given by
\vspace{-0.25cm}\begin{equation}
D_{min}(\theta_1,\theta_2) =(\theta_1-\theta_{12}^*)D_{1,min}+(\theta_2-\theta_{12}^*)D_{2,min}+(1+\theta_{12}^*-\theta_1-\theta_2)D_{max},\label{eqn:Dmin0}
\end{equation}
\vspace{-0.25cm}where $\theta_{12}^* = \theta_{12,max}$ if $D_{max}<D_{1,min}+D_{2,min}$, $\theta_{12}^* = \theta_{12,min}$ if $D_{max}>D_{1,min}+D_{2,min}$, and $\theta_{12}^*$ is arbitrary if $D_{max}=D_{1,min}+D_{2,min}$.
}\vspace{-0.25cm}\begin{proof}
It follows by considering the monotonicity of (\ref{eqn:Dmin_theta12}) with respect to $\theta_{12}$.
\end{proof}
\vspace{-0.25cm}{\lemma \label{lem:lem2} $D(R,\theta_{1},\theta_{2})$ is continuous and convex in $R$ for $R\geq0$. Similarly, $R(D,\theta_{1},\theta_{2})$ is continuous and convex in $D$ for $D\geq D_{min}(\theta_{1},\theta_{2})$. }
\vspace{-0.25cm}\begin{proof}It follows from the operation definitions given in Definition 1 similar to \cite[Thm. 10.2.1]{processingcost:Cover91}.
\end{proof}\vspace{-0.25cm}

\vspace{-0.25cm}
\section{Gaussian Sources}
\vspace{-0.15cm}
In this section, we focus on the case in which sources $S_{1}$ and
$S_{2}$ are jointly Gaussian, zero-mean, unit-variance and correlated
with coefficient $\rho$, with $\rho\in[-1,1]$. The decoder wishes
to compute a weighted sum function $T=w_{1}S_{1}+w_{2}S_{2}$,
with $w_{1},w_{2}\in\mathbb{R}$, under the mean square error (MSE)
distortion measure $d(t,\hat{t})=(t-\hat{t})^{2}$. In the following,
we study two specific choices for the weights $w_{1}=1,w_{2}=0$ and
$w_{1}=w_{2}=1$, resulting in the functions $T=S_{1}$
and $T=S_{1}+S_{2}$, respectively. These two cases are selected in
order to illustrate the impact of the choice of the function $f(S_{1},S_{2})$
on the optimal sampling strategy. The discussion can be extended with
appropriate modifications to arbitrary choices of weights $(w_{1},w_{2})$.

\vspace{-0.25cm}
\subsection{Computation of $T=S_{1}$} \label{subsec:cmpSrc1}
\vspace{-0.25cm}
{\proposition \label{prop:prop4} For a given sampling budget $(\theta_{1},\theta_{2})$, the distortion-rate function for
computing $T=S_{1}$ is
\vspace{-0.25cm}\begin{equation}
D(R,\theta_{1},\theta_{2})=\begin{cases}
1-\theta_{1}+\theta_{1}2^{-\frac{2R}{\theta_{1}}},\quad\text{if }R\leq\frac{\theta_{1}}{2}\log_{2}\left(\frac{1}{\rho^{2}}\right)\\
1-\theta_{1}-\rho^{2}(\theta_{2}-\theta_{12}^{*})+(\theta_{1}+\theta_{2}-\theta_{12}^{*})2^{-\frac{2R}{\theta_{1}+\theta_{2}-\theta_{12}^{*}}}\left(\rho^{2}\right)^{\frac{\theta_{2}-\theta_{12}^{*}}{\theta_{1}+\theta_{2}-\theta_{12}^{*}}},\;\text{otherwise }
\end{cases},\label{eqn:dist_rate_sum}
\end{equation}
\vspace{-0.25cm}where $\theta_{12}^{*}=\theta_{12,min}$ is the optimal overlap fraction. The
rate-distortion function $R(D,\theta_{1},\theta_{2})$ can be obtained
by inverting function (\ref{eqn:dist_rate_sum}) with respect to variable
$D$. } \begin{proof} See Appendix \ref{app:gaussian_sum}. \end{proof}

\vspace{-0.25cm}Proposition \ref{prop:prop4} confirms the intuition that if the receiver
is interested in source 1 only, i.e., $T=S_{1}$, the encoder should
simultaneously measure both sources $S_{1}$ and $S_{2}$ for
a fraction of time to be kept as small as possible. Moreover, if $R\leq(\theta_{1}/2)\log_{2}(1/\rho^{2})$,
the entire rate $R$ is used to describe only the $\theta_{1}$-fraction
of samples measured from source $S_{1}$ only; otherwise, both
the $\theta_{1}$-fraction of source $S_{1}$ and the $(\theta_{2}-\theta_{12}^{*})$-fraction
of source $S_{2}$ that is not overlapped are described at positive
rates. Using a variant of the classic reverse water-filling solution \cite{processingcost:Cover91}, the threshold value of rate $R$, for which only the independent source with the larger variance, namely, the $\theta_1$-fraction, is described, can be obtained as $(\theta_1/2)\log_2(1/\rho^2)$. This threshold only depends on the sampling fraction of the source with the larger variance, $\theta_1$, and the ratio of the variances of the $\theta_1$-fraction and the $(\theta_2-\theta_{12}^*)$-fraction, $1/\rho^2$. The reader is referred to Appendix A and \cite{processingcost:Liu_Simeone_Erkip_tcom} for more details on how rate R is
optimally allocated between the two fractions of source samples.
\vspace{-0.25cm}
\subsection{Computation of $T=S_{1}+S_{2}$}

We now consider the case in which the desired function is $T=S_{1}+S_{2}$.
Note that $T$ is a Gaussian random variable with zero mean and variance
$D_{max}=2(1+\rho)$, and that $T$ and $S_{1}$ (or $S_{2}$) are
jointly Gaussian with correlation coefficient $\tilde{\rho}=\sqrt{(1+\rho)/2}$.
Moreover, since $T=0$ for $\rho=-1$, it is enough to focus on the interval $\rho\in(-1,1]$.
Finally, we recall that the distortion-rate function for $T$
is given by $D_{12}(R)=2(1+\rho)2^{-2R}$ for $R\geq0$ \cite{processingcost:Cover91},
and the indirect distortion-rate function is $D_{k}(R)=2(1+\rho)(1-\tilde{\rho}^{2}+\tilde{\rho}^{2}2^{-2R})$
 for $R\geq0$ and $k=1,2$ \cite{processingcost:Oohama_IT97}.

{\proposition \label{prop:prop6} Given sampling budget
$(\theta_{1},\theta_{2})$, the distortion-rate function for computing
$T=S_{1}+S_{2}$ is
\begin{align}
 D(R,\theta_{1},\theta_{2})=&\min_{\theta_{12},R_{12}}(1+\rho)^{2}(\theta_{1}+\theta_{2}-2\theta_{12})2^{-\frac{2(R-R_{12})}{\theta_{1}+\theta_{2}-2\theta_{12}}}\nonumber \\
 &+2(1+\rho)\left(1+\rho\theta_{12}+\theta_{12}2^{-\frac{2R_{12}}{\theta_{12}}}\right)-(1+\rho)^{2}(\theta_{1}+\theta_{2}),\label{eqn:opt2}
\end{align}
 where the minimization is taken over all $\theta_{12}$
satisfying (\ref{eqn:cnst_theta12}) and all $R_{12}$ satisfying
$R_{12}\leq R$. }
\vspace{-0.15cm}\begin{proof}This proposition follows by Lemma \ref{lem:lem1} using arguments similar to the ones in Appendix \ref{app:gaussian_sum}. \end{proof}

\vspace{-0.15cm}In order to obtain further analytical
insight into (\ref{eqn:opt2}) and the optimal sampling strategy, we now consider some special
cases of interest.

\vspace{-0.15cm}{\proposition \label{cor:cor1} For $R\rightarrow\infty$, we have
\begin{equation}
D_{min}(\theta_{1},\theta_{2})=2(1+\rho)(1+\rho\theta_{12}^{*})-(1+\rho)^{2}(\theta_{1}+\theta_{2}),
\end{equation}
\vspace{-0.15cm}where $\theta_{12}^{*}=\theta_{12,min}$ if $\rho>0$, $\theta_{12}^{*}=\theta_{12,max}$
if $\rho<0$, and $\theta_{12}^{*}$ is arbitrary if $\rho=0$. }

This proposition is easily obtained from Lemma \ref{lem:lem_Dmin}. It says that,
if the sources ($S_{1},S_{2}$) have positive correlation, i.e., $\rho>0$,
and there are no rate limitations ($R\rightarrow\infty$), the MSE
distortion increases linearly with $\theta_{12}$, and it is thus
optimal to set $\theta_{12}$ to be the smallest possible value $\theta_{12}^{*}=\theta_{12,min}$.
In contrast, if $\rho<0$, the MSE distortion decreases linearly with
$\theta_{12}$, and thus the optimal $\theta_{12}^{*}$ is the largest
possible value, $\theta_{12}^{*}=\theta_{12,max}$. This shows the
relevance of the source correlation in designing the optimal sampling
strategy.

The general conclusions about the optimal sampling strategies discussed
above for infinite rate can be extended to finite rates $R$ in certain regimes.
Specifically, Proposition \ref{cor:cor2} below states that if $\rho\leq0$,
then, just as in the case of $R\rightarrow \infty$ of Proposition \ref{cor:cor1},
the encoder should set $\theta_{12}$ to be as large as possible,
i.e., $\theta_{12}^{*}=\theta_{12,max}$, irrespective of the value of $R$.
Furthermore, Proposition \ref{cor:cor3} below demonstrates that, for sufficiently small rates,
the optimal overlap $\theta_{12}^{*}$
tends to be maximum, i.e., $\theta_{12}^{*}=\theta_{12,max}$, for
a larger range of correlation coefficients $\rho$ than $\rho\leq 0$. This is mainly because when rate $R$ is small enough,
it is generally more efficient to use the available rate to describe
$T$ directly during the overlapping $\theta_{12}$-fraction,
rather than indirectly describing $T$ based on observations of
$S_{1}$ or $S_{2}$ alone.

\vspace{-0.15cm}{\proposition \label{cor:cor2} For $\rho\leq0$, the distortion-rate
function is
\begin{equation}\vspace{-0.1cm}
D(R,\theta_{1},\theta_{2})=\begin{cases}
(\theta_{1}+\theta_{2}-\theta_{12}^{*})(1+\rho)^{2}2^{-\frac{2R}{\theta_{1}+\theta_{2}-\theta_{12}^{*}}}\left(\frac{2}{1+\rho}\right)^{\frac{\theta_{12}^{*}}{\theta_{1}+\theta_{2}-\theta_{12}^{*}}}\\
\hspace{0.25cm}+2(1+\rho)(1+\rho\theta_{12}^{*})-(1+\rho)^{2}(\theta_{1}+\theta_{2}),\quad \text{if } R>\frac{\theta_{12}^{*}}{2}\log_{2}\left(\frac{2}{1+\rho}\right),\\
2(1+\rho)\left(1-\theta_{12}^{*}+\theta_{12}^{*}2^{-\frac{2R}{\theta_{12}^{*}}}\right),\quad\text{otherwise},
\end{cases}\vspace{-0.1cm}
\end{equation}
\vspace{-0.15cm}where $\theta_{12}^{*}=\theta_{12,max}$ is the optimal overlapping
fraction. } \vspace{-0.1cm}\begin{proof} The proof is obtained by solving (\ref{eqn:opt2}) for $\rho\leq 0$. \end{proof}

\vspace{-0.15cm}{\proposition \label{cor:cor3} For any $\rho>0$, if $R\leq(\theta_{12,min}/2)\log_{2}(2/(1+\rho))$,
the distortion-rate function is given as
\begin{equation}\vspace{-0.1cm}
D(R,\theta_{1},\theta_{2})=2(1+\rho)(1-\theta_{12}^{*})+2(1+\rho)\theta_{12}^{*}2^{-\frac{2R}{\theta_{12}^{*}}},
\vspace{-0.1cm}\end{equation}
\vspace{-0.25cm}where $\theta_{12}^{*}=\theta_{12,max}$.  }\begin{proof} Given $R\leq(\theta_{12,min}/2)\log_2(2/(1+\rho))$, for any feasible $\theta_{12}$ satisfying (\ref{eqn:cnst_theta12}), we always have $R\leq (\theta_{12}/2)\log_2(2/(1+\rho))$. In this case, for any given $\theta_{12}$, applying the standard Lagrangian method to (\ref{eqn:opt2}), we obtain $R_{12}^* = R$. Substituting into (\ref{eqn:opt2}) and considering the monotonicity of function $D(R,\theta_1,\theta_2,\theta_{12})$ with respect to $\theta_{12}$, we can show that the optimal overlap fraction is given by $\theta_{12}^* = \theta_{12,max}$, leading to the distortion-rate function as stated in the proposition.\end{proof}

\vspace{-0.25cm}\subsection{Numerical Results}

\begin{figure}
\centering \includegraphics[width=2.9in]{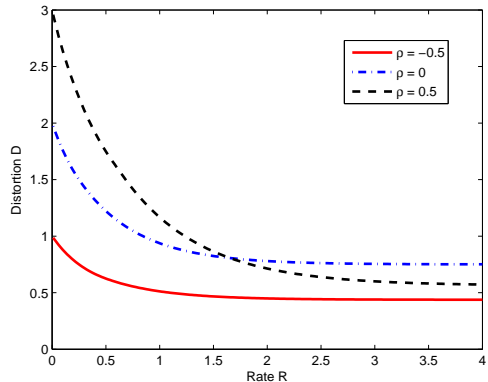} \caption{Distortion-rate function for computing $T = S_1+S_2$, $(S_1, S_2)$ jointly Gaussian with correlation coefficient $\rho=-0.5,0,0.5$, respectively. The sampling budget is $(\theta_{1},\theta_{2})=(0.5,0.75)$.}
\label{fig:dist_vs_rate}
\end{figure}

\begin{figure}
\centering \includegraphics[width=2.9in]{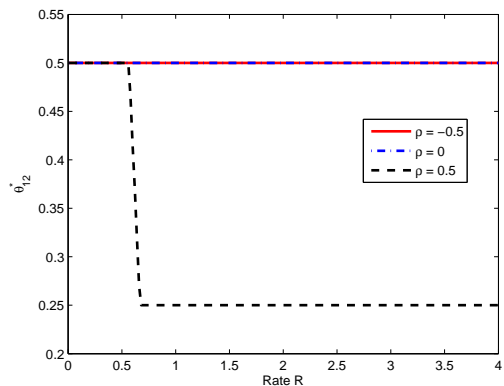} \caption{Optimal overlap fraction $\theta_{12}^{*}$ that minimizes the average expected distortion as a function of rate
$R$ for computing $T = S_1+S_2$, $(S_1, S_2)$ jointly Gaussian with correlation coefficient $\rho=-0.5,0,0.5$, respectively. The sampling budget is $(\theta_{1},\theta_{2})=(0.5,0.75)$.}
\label{fig:opt_theta12}
\end{figure}

In this subsection, we numerically evaluate the distortion-rate function with fractional sampling for computation of function $T = S_1+S_2$. Fig. \ref{fig:dist_vs_rate} and Fig. \ref{fig:opt_theta12} show
the minimum MSE distortion $D$ and the optimal overlap fraction $\theta_{12}^{*}$
versus rate $R$, respectively, for $(\theta_{1},\theta_{2})=(0.5,0.75)$,
and $\rho=-0.5,0,0.5$. The curves are obtained by numerically solving
the optimization in (\ref{eqn:opt2}). It can be seen from Fig. \ref{fig:opt_theta12}
that, as predicted by Proposition \ref{cor:cor2}, the optimal overlap
fraction $\theta_{12}^{*}$ is equal to the maximum possible fraction
$\theta_{12,max}=0.5$, for $\rho=-0.5<0$ and $\rho=0$. Moreover,
for $\rho=0.5>0$, with sufficiently small rates $R$, as described
in Proposition \ref{cor:cor3}, the optimal overlap fraction $\theta_{12}^{*}$
equals to the maximum overlap $\theta_{12,max}=0.5$. However, as
$R$ increases, $\theta_{12}^{*}$ drops to the
minimum value $\theta_{12,min}=0.25$, which is consistent with Proposition
\ref{cor:cor1}.
\vspace{-0.25cm}

\section{Binary Sources}

In this section, we consider binary sources so that $\mathcal{S}_{1}=\mathcal{S}_{2}=\mathcal{T}=\hat{\mathcal{T}}=\{0,1\}$,
and $(S_{1},S_{2})$ is a doubly symmetric binary source (DSBS), i.e., we have $S_1\sim \text{Bernoulli}(1/2)$, $S_2\sim \text{Bernoulli}(1/2)$ and $S_1 \oplus S_2 \sim \text{Bernoulli}(p)$, where $0\leq p\leq1/2$. In other words, $S_2$ is the output of a binary symmetric channel with crossover probability $p$ corresponding to the input $S_1$. We take the Hamming distortion as the distortion measure, i.e., $d(t,\hat{t})=1-\delta_{t\hat{t}}$,
where $\delta_{t\hat{t}}=1$ if $t=\hat{t}$ and $\delta_{t\hat{t}}=0$
otherwise. Since all non-trivial binary functions are equivalent,
up to relabeling, to either the exclusive OR or the AND \cite{processingcost:Yamamoto82},
it suffices to consider only these two options for function $T=f(S_{1},S_{2})$:
(\emph{i}) the exclusive OR or binary sum, i.e., $T=S_{1}\oplus S_{2}$;
(\emph{ii}) the AND or binary product, i.e., $T=S_{1}\otimes S_{2}$.
In the following, we focus on deriving the rate-distortion $R(D,\theta_{1},\theta_{2})$
for convenience, since in general it takes a simpler analytical form
as compared to the distortion-rate function $D(R,\theta_{1},\theta_{2})$.

\vspace{-0.25cm}\subsection{Computation of $T=S_{1}\oplus S_{2}$}

\vspace{-0.25cm}{\proposition For given sampling budget $(\theta_{1},\theta_{2})$,
the rate-distortion function for computing $T=S_{1}\oplus S_{2}$
is given by
\vspace{-0.25cm}\begin{equation}
R(D,\theta_{1},\theta_{2})=\begin{cases}
h(p)-h\left(\frac{D-(1-\theta_{12}^{*})p}{\theta_{12}^{*}}\right),\quad\text{if }(1-\theta_{12}^{*})p\leq D<p,\\
0,\quad\quad\quad\text{if }D\geq p,
\end{cases}\label{eqn:rate_dist_binsum}
\end{equation}
\vspace{-0.25cm}where $h(x)=-x\log_{2}(x)-(1-x)\log_{2}(1-x)$ is the binary entropy
function, and $\theta_{12}^{*}=\theta_{12,max}$ is the optimal overlap
fraction, for $(1-\theta_{12}^{*})p\leq D<p$. }

\vspace{-0.25cm}
\begin{proof}
Since $T=S_{1}\oplus S_{2}$ is a Bernoulli($p$) random variable independent of $S_{1}$
and $S_{2}$, the observation of either $S_{1}$ or $S_{2}$
is not useful for computing $T$. Thus, one should choose the overlap
fraction to be as large as possible, i.e., $\theta_{12}^{*}=\theta_{12,max}$.
The rate-distortion function (\ref{eqn:rate_dist_binsum}) then follows
immediately from the rate-distortion function of the binary random
variable $T$ \cite{processingcost:Cover91}.\end{proof}

\vspace{-0.25cm}
\subsection{Computation of $T=S_{1}\otimes S_{2}$}

In this subsection, we focus on the binary product $T=S_{1}\otimes S_{2}$,
which is Bernoulli distributed with probability $(1-p)/2$. For convenience,
we start by finding the minimum possible distortion at the decoder
given $(\theta_{1},\theta_{2})$, i.e., $D_{min}(\theta_{1},\theta_{2})$ as defined in Lemma \ref{lem:lem2},
and the minimum required rate to achieve it. Then, we proceed to derive
the rate-distortion function.
{\proposition \label{prop:prop3}
For given sampling budget $(\theta_{1},\theta_{2})$, the
minimum achievable distortion for computing $T=S_{1}\otimes S_{2}$
is given by
\begin{equation}
D_{min}(\theta_{1},\theta_{2})=\frac{1-p}{2}+\left(p-\frac{1}{2}\right)(\theta_{1}+\theta_{2})+\left(\frac{1-3p}{2}\right)\theta_{12}^{*},
\end{equation}
 where $\theta_{12}^{*}=\theta_{12,min}$ if $p<1/3$ and $\theta_{12}^{*}=\theta_{12,max}$ if $1/3\leq p\leq1/2$. Moreover, distortion
$D_{min}(\theta_{1},\theta_{2})$ can be achieved as long as $R\geq R_{min}(\theta_1,\theta_2)=\theta_{1}+\theta_{2}-\left(2-h\left(\frac{1-p}{2}\right)\right)\theta_{12}^{*}$.}
\begin{proof}See Appendix \ref{app:bin_prod_largeR}.\end{proof}

The results in Proposition \ref{prop:prop3} can be seen as the counterpart
of Proposition \ref{cor:cor1} for binary sources. In fact, they show
that, for sufficiently large $R$, if $p<1/3$, the average
Hamming distortion increases linearly with $\theta_{12}$ and thus
we should set $\theta_{12}$ to the smallest possible value $\theta_{12,min}$;
instead, if $1/3\leq p\leq1/2$, the optimal value of $\theta_{12}$
is the largest possible, namely, $\theta_{12,max}$.

Before we proceed to investigate the general rate-distortion function
$R(D,\theta_{1},\theta_{2})$, we first derive the indirect rate-distortion
function $R_{1}(D)$ when only $S_{1}$ is observed at the encoder.
\vspace{-0.25cm}{\lemma \label{lem:lem3} The indirect
rate-distortion function for $T=S_{1}\otimes S_{2}$ is given by
\begin{align}
R_{1}(D)=\begin{cases}
{\displaystyle \min_{\frac{1-p-2D}{1-2p}\leq y\leq1}h\left(D+y(1-p)+\frac{p-1}{2}\right)-\frac{1}{2}h(y)}\\
\hspace{0.4cm}-\frac{1}{2}h(2D+y(1-2p)+p-1),\quad\frac{p}{2}<D\leq\frac{1-p}{2},\\
0,\quad D\geq\frac{1-p}{2},
\end{cases}\label{eqn:R1_D}
\end{align}
}\begin{proof}See Appendix \ref{app:app1}.\end{proof}

By symmetry, the indirect rate-distortion function $R_{2}(D)$ for
$T$ when $S_{2}$ is observed at the encoder is also given by Lemma
\ref{lem:lem3}. The rate-distortion function $R_{12}(D)$ for
$T$ is obtained from standard results \cite{processingcost:Cover91}
as $R_{12}(D)=h((1-p)/2)-h(D)$ if $0\leq D\leq(1-p)/2$, and $R_{12}(D)=0$
if $D>(1-p)/2$.

{\proposition \label{prop:prop5} For a given sampling budget $(\theta_{1},\theta_{2})$,
the rate-distortion function for computing $T=S_{1}\otimes S_{2}$
is given as
\begin{align}
R(D,\theta_{1},\theta_{2})=\begin{cases}
{\displaystyle \min_{\theta_{12},D_{3},D_{12}}\theta_{12}\left(h\left(\frac{1-p}{2}\right)-h\left(\frac{D_{12}}{\theta_{12}}\right)\right)}+(\theta_{1}+\theta_{2}-2\theta_{12})R_{1}\left(\frac{D_{3}}{\theta_{1}+\theta_{2}-2\theta_{12}}\right),\\
\hspace{2.1cm}\text{if }D_{min}(\theta_{1},\theta_{2})\leq D<\frac{1-p}{2},\\
0,\hspace{1.7cm}\text{if }D\geq\frac{1-p}{2},
\end{cases}
\end{align}
 where $D_{min}(\theta_{1},\theta_{2})$ is as given in Proposition \ref{prop:prop3} and the
minimization is taken over all choices of $\theta_{12}$, $D_{3}$
and $D_{12}$ such that (\ref{eqn:cnst_theta12}) is satisfied, $p(\theta_{1}+\theta_{2}-2\theta_{12})/2\leq D_{3}\leq(1-p)(\theta_{1}+\theta_{2}-2\theta_{12})/2$,
$p\theta_{12}/2\leq D_{12}\leq(1-p)\theta_{12}/2$, and
\begin{equation}
D_{3}+D_{12}+\left(\frac{1-p}{2}\right)(1+\theta_{12}-\theta_{1}-\theta_{2})=D.\label{eqn:dist_sum_equality}
\end{equation}
 }\begin{proof}See Appendix \ref{app:bin_prod}.\end{proof}

\subsection{Numerical Results}

\begin{figure}
\centering
\includegraphics[width = 3in]{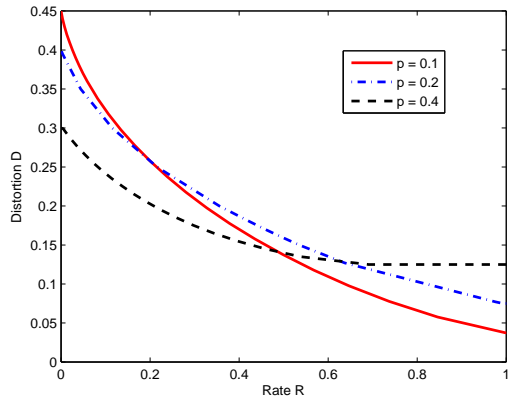}
\caption{Distortion-rate function for computing $T = S_1\otimes S_2$, $(S_1,S_2)$ doubly symmetric binary with probability $\text{Pr}[S_{1}\neq S_{2}]$ equal to $p = 0.1,0.2,0.4$, respectively. The sampling budget is $(\theta_{1},\theta_{2})=(0.5,0.75)$.}
\label{fig:dist_vs_rate_bin}\vspace{-0.5cm}
\end{figure}

\begin{figure}
\centering
\includegraphics[width = 3in]{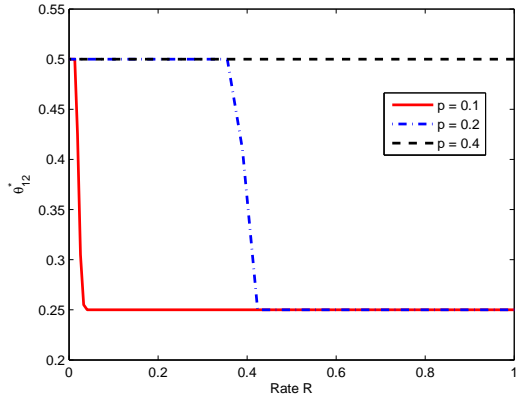}
\caption{Optimal overlap fraction $\theta_{12}^*$ that minimizes the average expected distortion as a function of $R$ for computing $T = S_1\otimes S_2$, $(S_1,S_2)$ doubly symmetric binary with probability $\text{Pr}[S_{1}\neq S_{2}]$ equal to $p = 0.1,0.2,0.4$, respectively. The sampling budget is $(\theta_{1},\theta_{2})=(0.5,0.75)$.}
\label{fig:opt_theta12_bin}
\end{figure}

In this subsection, we numerically evaluate the distortion-rate function for computation of function $T = S_1\otimes S_2$. Fig. \ref{fig:dist_vs_rate_bin} and Fig. \ref{fig:opt_theta12_bin} plot the minimum average Hamming distortion $D$ and the optimal overlap fraction $\theta_{12}^*$ for $(\theta_1,\theta_2) = (0.5, 0.75)$, and $p = 0.1, 0.2, 0.4$. In Fig. \ref{fig:dist_vs_rate_bin}, as predicted by Proposition \ref{prop:prop3}, the minimum rate $R_{min}(\theta_1,\theta_2)$ that achieves distortion $D_{min}(\theta_{1},\theta_{2})$, is given by $0.9982$, $0.9927$, $0.69$ for $p = 0.1, 0.2, 0.4$, respectively. It can be observed from Fig. \ref{fig:opt_theta12_bin}, for $p = 0.4>1/3$, the optimal overlap fraction $\theta_{12}^*$ is equal to the maximum possible value $\theta_{12,max} = 0.5$, for any $0\leq R\leq 1$. However, for smaller probabilities $p = 0.1,0.2$, the optimal overlap fraction equals to the maximum possible value $\theta_{12,max} = 0.5$ for sufficiently smaller rates and then drops to the minimum possible value $\theta_{12,min} = 0.25$ once $R$ gets larger. Moreover, the smaller the probability $p$ is, the larger range of rates $R$ over which the optimal overlap fraction $\theta_{12}^*$ is $\theta_{12,min} = 0.25$.
We note that with a larger $p$, it is easier to describe $T$ directly, since $T\sim \text{Bernoulli}((1-p)/2)$, but the indirect description of $T$ based on $S_1$ or $S_2$ becomes more difficult since $T$ becomes less correlated with $S_1$ or $S_2$.\footnote{The correlation coefficient between $T$ and $S_1$ or $S_2$ is given by $\sqrt{(1-p)/(1+p)}$.} This explains why the optimal overlap fraction should be chosen as the maximum possible value $\theta_{12,max} = 0.5$ when $p$ is larger than $1/3$ (see the curve $p = 0.4$). In this sense, the regime $p\geq 1/3$ may be considered as the binary counterpart of the regime $\rho\leq 0$ for the Gaussian sum case in Section IV-B. For probabilities $p<1/3$, the numerical results above imply that the optimal overlap depends on the link rate $R$. Similar to the Gaussian sum case when $\rho>0$ (Proposition \ref{cor:cor3}), when $R$ is sufficiently small, it remains optimal to choose the overlap fraction to be the maximum possible; however, as $R$ grows sufficiently large, it is more advantageous to have the overlap fraction as small as possible, which is consistent with Proposition \ref{prop:prop3}.

\vspace{-0.25cm}

\section{Multi-Hop Lossy Computing with Fractional Sampling}\label{sec:multi-hop}

In this section, we extend the analysis of lossy computing with fractional sampling from the point-to-point setup to a multi-hop setup as depicted in Fig. \ref{fig:Cascademodel}. We assume that Encoder $k$ can only sample a fraction $\theta_k$ of the samples for source $S_k$, with $0\leq \theta_k\leq 1$, for $k=1,2$. Moreover, the encoders make decisions on which samples to sense independently and based only on the statistics of the sources. In particular, to ensure causality, Encoder 2 is not allowed to observe the message from Encoder 1 before making a decision on which samples to measure as instead assumed in \cite{processingcost:Permuter_Weissman_IT11} for a related set-up. Under this assumption, the sampling fractions can overlap for a fraction $\theta_{12}$, with $\theta_{12}$ satisfying (\ref{eqn:cnst_theta12}). Similar to the point-to-point setup of Fig. \ref{fig:model}, it is without loss of generality to assume the sampling profile is as shown in Fig. \ref{fig:fraction}. The links between Encoder 1 and Encoder 2 and between Encoder 2 and the decoder can support a rate of $R_1$ bits/sample and a rate of $R_2$ bits/sample, respectively. As above, the goal is to estimate a function $T^n = f^n(S_1^n,S_2^n)$ at the decoder. It is observed that if $R_1$ is unbounded, then the scenario reduces to the point-to-point system studied in the previous sections.

\textit{Definition 5}: A $(n, R_1, R_2, D, \theta_1,\theta_2,\theta_{12})$ code for the problem of multi-hop lossy computing of two memoryless sources with fractional sampling consists of an encoder (Encoder 1) $f_1: \mathcal{S}_1^{n\theta_1}\rightarrow \{1,...,2^{nR_1}\}$; an encoder (Encoder 2) $
f_2: \mathcal{S}_2^{n\theta_2}\times\{1,...,2^{nR_1}\}\rightarrow\{1,...,2^{nR_2}\}$; and a decoder $g:\{1,...,2^{nR_2}\}\rightarrow \hat{\mathcal{T}}^n$ such that distortion constraint $D$ is satisfied as in (\ref{eqn:AvgDist0}). It is assumed that encoder $f_1$ operates on the measurements $(S_{1,1},...,S_{1,n\theta_1})$ and encoder $f_2$ operates on the measurements $(S_{2,n(\theta_1-\theta_{12})+1},...,S_{2,n(\theta_1+\theta_2-\theta_{12})})$ as well as the index received from Encoder 1.

The distortion-rate function $D(R_1,R_2,\theta_1,\theta_2,\theta_{12})$ and the distortion-rate function $D(R_1,R_2,\theta_1,\theta_2)$ are defined in a similar manner as in the point-to-point setup of Section II. In the remaining of this section, we focus on the specific case in which sources $S_1$ and $S_2$ are Gaussian and the decoder wishes to compute the sum $T = S_1+S_2$. Other cases studied in Section IV and Section V can also be investigated similarly.

\subsection{Lower Bounds on the Achievable Distortion}
Two lower bounds on the achievable distortion for the Gaussian multi-hop lossy computing problem discussed above can be derived based on the cut-set arguments \cite{processingcost:Gastpar}. Specifically, the first cut is around Encoder 1 and the second cut is around the decoder. These two cuts induce the following two subproblems of the original problem in Fig. \ref{fig:Cascademodel}: 1) For the cut around Encoder 1, the problem is equivalent to point-to-point lossy computing with side information, in which the encoder and the decoder can only measure a fraction of the $n$ samples from the respective source; 2) For the cut around the decoder, the problem reduces to that of point-to-point source coding problem investigated in Section IV-B, leading to a lower bound as given by (\ref{eqn:opt2}) of Proposition \ref{prop:prop6} with $R$ replaced by $R_2$.

\begin{figure}[ht]
\centering
\includegraphics[scale=0.5]{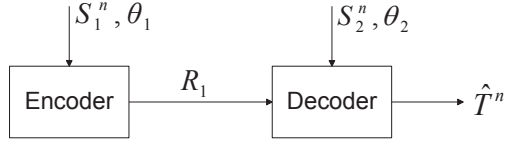}
\caption{Lossy computing with side information at the decoder and fractional sampling: The encoder measures source $S_1$ for a fraction  of time $\theta_1$, and the decoder measures source $S_2$ for a fraction of time $\theta_2$ and estimates a function $T^n = f^n(S_1^n,S_2^n)$. }\label{fig:p2p1}
\vspace{-0.5cm}
\end{figure}

In the following, we study the first subproblem identified above, namely, the problem of lossy computing with side information at the decoder and fractional sampling as shown in Fig. \ref{fig:p2p1}. To elaborate, the encoder measures a fraction $\theta_1$ of samples from $S_1$ and describes it using rate $R_1$ to the decoder. At the same time, the decoder measures a fraction, $\theta_2$, of samples of a correlated source $S_2$, which overlaps with the encoder's measurements for a fraction $\theta_{12}$ of samples. Based on the description received from the encoder and its own measurements, the decoder forms the estimate $\hat{T}^n$.

{\proposition \label{prop:lb1} For the problem of lossy computing with side information at the decoder and fractional sampling, given sampling budget $(\theta_1,\theta_2)$ and rate $R_1$, the distortion-rate function $D^{(1)}(R_1,\theta_1,\theta_2)$ can be obtained as follows.
\begin{enumerate}
\item For $\rho>0$,
\begin{equation}
D^{(1)}(R_1,\theta_1,\theta_2)=  \begin{cases}
(1+\rho)^2(\theta_1-\theta_{12}^*)2^{-\frac{2R_1}{\theta_1-\theta_{12}^*}}+(1+\rho)^2(\theta_{12}^*-\theta_1-\theta_2)+2(1+\rho), \\
\hspace{5cm}\text{if } R_1\leq \frac{\theta_1-\theta_{12}^*}{2}\log_2\left(\frac{1+\rho}{1-\rho}\right),  \\
\theta_1 (1+\rho)^2\left(\frac{1-\rho}{1+\rho}\right)^{\frac{\theta_{12}^*}{\theta_1}} 2^{-\frac{2R_1}{\theta_1}} +2\rho(1+\rho)\theta_{12}^*- (1+\rho)^2(\theta_1+\theta_2)+2(1+\rho),\\
\hspace{5cm}\text{otherwise,}
\end{cases}
\label{eqn:lowerbound1}
\end{equation}
where $\theta_{12}^* = \theta_{12,min}$ is the optimal overlap fraction;
\item If $\rho\leq 0$, then
\begin{equation}
D^{(1)} (R_1,\theta_1,\theta_2) =  \begin{cases}
(1-\rho^2)\theta_{12}^*2^{-\frac{2R_1}{\theta_{12}^*}} -(1-\rho^2)\theta_{12}^* -(1+\rho)^2\theta_2+2(1+\rho),\\
\hspace{5cm}\text{if } R_1\leq \frac{\theta_{12}^*}{2}\log_2\left(\frac{1-\rho}{1+\rho}\right),  \\
\theta_1 (1+\rho)^2\left(\frac{1-\rho}{1+\rho}\right)^{\frac{\theta_{12}^*}{\theta_1}} 2^{-\frac{2R_1}{\theta_1}} +2\rho(1+\rho)\theta_{12}^*- (1+\rho)^2(\theta_1+\theta_2)+2(1+\rho),\\
\hspace{5cm}\text{otherwise,}
\end{cases}
\label{eqn:lowerbound1_2}
\end{equation}
where $\theta_{12}^* = \theta_{12,max}$ is the optimal overlap fraction.
\end{enumerate}
}
\begin{proof}
See Appendix \ref{app:lowerbound1}.
\end{proof}

Proposition \ref{prop:lb1} states that, in the setting of Fig. \ref{fig:p2p1}, it is optimal to have the overlap fraction as small as possible if the two sources are positively correlated, and as large as possible if they are negatively correlated. This result is consistent and follows from similar considerations as Proposition \ref{cor:cor1}, \ref{cor:cor2} and \ref{cor:cor3}.

\subsection{Upper Bounds on the Achievable Distortion}
In this subsection, we first propose a specific strategy, thus providing an upper bound on the achievable distortion.  The derived upper bound is then compared to the lower bounds in Proposition \ref{prop:prop6} and Proposition \ref{prop:lb1} through numerical examples.

In the proposed strategy, we treat the $(\theta_1-\theta_{12})$-fraction of samples measured only at Encoder 1, the overlapping $\theta_{12}$-fraction measured by both the encoders, and the $(\theta_2-\theta_{12})$-fraction measured only at Encoder 2 separately in terms of encoding and decoding. In particular, on the link between the two encoders, which is of rate $R_1$, we assign rate $R_{11}$ to the encoded version of the $(\theta_1-\theta_{12})$-fraction of samples and rate $R_{12}$ to the $\theta_{12}$-fraction. Moreover, on the link between Encoder 2 and the decoder, which is of rate $R_2$, we allocate rate $R_{21}$ to forward the encoded version of the $(\theta_1-\theta_{12})$-fraction, rate $R_{22}$ to the $\theta_{12}$-fraction and rate $R_{23}$ to the $(\theta_2-\theta_{12})$-fraction. By definition, we thus have the conditions
\vspace{-0.25cm}\begin{subequations}\begin{align}
&R_{11}+R_{12}\leq R_1,\\
\text{and }& R_{21}+R_{22}+R_{23}\leq R_2.
\end{align}\label{eqn:cnst_r1r2}\vspace{-0.15cm}\end{subequations}
We specify the source coding strategy used for the different fractions of samples and discuss the resulting average distortions as follows. For the $(\theta_1-\theta_{12})$-fraction measured only by Encoder 1, we have available an end-to-end rate equal to $\min(R_{11},R_{21})$. Encoder 1 thus compresses this fraction of samples of $S_1$ at rate $\min(R_{11},R_{21})/(\theta_1-\theta_{12})$ bits/source sample using a standard indirect rate-distortion optimal code, leading to average distortion $1-\rho^2+(1+\rho)^2 2^{-\frac{2\min(R_{11},R_{21})}{\theta_1-\theta_{12}}}$ \cite{processingcost:Oohama_IT97}. Similarly, for the $(\theta_2-\theta_{12})$-fraction of samples measured only by Encoder 2, Encoder 2 can employ rate $R_{23}/(\theta_2-\theta_{12})$  using a standard indirect rate-distortion optimal code, leading to average distortion $1-\rho^2 + (1+\rho)^2 2^{-\frac{2R_{23}}{\theta_2-\theta_{12}}}$ \cite{processingcost:Oohama_IT97}. For the $(1+\theta_{12}-\theta_1-\theta_2)$-fraction measured by neither node, the average distortion at the decoder is equal to the variance of source $T = S_1+S_2$, namely, $2(1+\rho)$. For the $\theta_{12}$-fraction measured by both nodes, the setup at hand reduces to the multi-hop source coding problem investigated in \cite{processingcost:Cuff_Su_Gamal09} with the average link rates over the two links being $R_{12}/\theta_{12}$ and $R_{22}/\theta_{12}$, respectively. Among the class of achievable schemes considered in \cite{processingcost:Cuff_Su_Gamal09}, under the assumption of unit-variance sources, the so called ``re-compress''
scheme is optimal. Therefore, we assume the ``re-compress'' scheme for the $\theta_{12}$-fraction at hand, which leads to  average distortion $D_0(R_{12}/\theta_{12},R_{22}/\theta_{12})$, where $D_0(R_1,R_2) = (1-\rho^2)(1-2^{-2R_2})2^{-2R_1}+2(1+\rho)2^{-2R_2}$, for $R_1\geq 0$ and $R_2\geq 0$.

Applying the source coding strategy described above, an achievable distortion at the decoder is given by summing the contributions of the different fractions of samples with the appropriate weights as
\vspace{-0.15cm}\begin{align}
D_{ub}(R_1,R_2,\theta_1,\theta_2) = &\min_{\theta_{12},R_{11}, R_{22}} (\theta_1-\theta_{12})(1+\rho)^2 2^{-\frac{2R_{11}}{\theta_1-\theta_{12}}}  + \theta_{12} D_0\left(\frac{R_1-R_{11}}{\theta_{12}},\frac{R_{22}}{\theta_{12}}\right)+ (1+\rho)^2\nonumber\\
 & \cdot(\theta_2-\theta_{12})2^{-\frac{2(R_2-R_{11}-R_{22})}{\theta_2-\theta_{12}}} + 2\rho(1+\rho)\theta_{12}-(1+\rho)^2(\theta_1+\theta_2)+2(1+\rho) ,\label{eqn:d_up}
\end{align}
\vspace{-0.15cm}where the minimum is taken under the constraints (1) and (\ref{eqn:cnst_r1r2}). Note that in (\ref{eqn:d_up}), we have set $R_{12} = R_1-R_{11}$, $R_{21} = R_{11}$ and $R_{23} = R_2 - R_{11}-R_{22}$ without loss of optimality, since the two rate bounds in (\ref{eqn:cnst_r1r2}) are easily seen to be satisfied with equality at an optimal solution.

\subsection{Numerical Results}
\begin{figure}
\centering
\includegraphics[width = 3in]{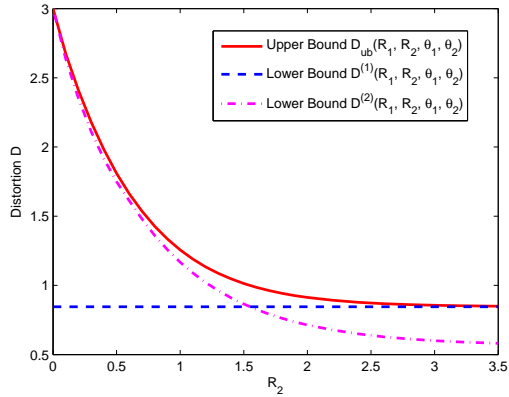}
\caption{The upper and lower bounds on the achievable average distortion versus link rate $R_2$ for computing $T=S_1+S_2$, $(S_1, S_2)$ jointly Gaussian with correlation coefficient $\rho = 0.5$. The sampling budget is $(\theta_1, \theta_2) = (0.5,0.75)$ and the link rate between the encoders is $R_1 = 0.3$.}
\label{fig:mh1}
\end{figure}

\begin{figure}
\centering
\includegraphics[width = 3in]{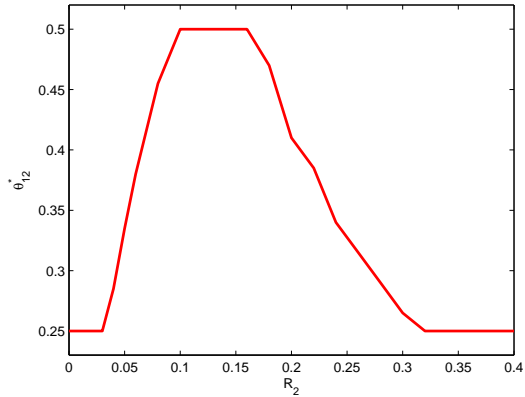}
\caption{The optimal overlap fraction $\theta_{12}^*$ that minimizes the achievable average distortion versus link rate $R_2$ for computing $T=S_1+S_2$, $(S_1,S_2)$ jointly Gaussian with correlation coefficient $\rho = 0.5$. The sampling budget is $(\theta_1, \theta_2) = (0.5,0.75)$ and the link rate between the encoders is $R_1 = 0.3$.}
\label{fig:mh2}
\end{figure}

Fig. \ref{fig:mh1} and Fig. \ref{fig:mh2} plot the achievable distortion $D_{ub}(R_1,R_2,\theta_1,\theta_2)$ and the corresponding optimized overlap fraction $\theta_{12}^*$ as a function of rate $R_2$ when $R_1 = 0.3$, $(\theta_1, \theta_2) = (0.5,0.75)$ and $\rho = 0.5$. In Fig. \ref{fig:mh1}, the two lower bounds obtained, $D^{(1)}(R_1,R_2,\theta_1,\theta_2)$ from Proposition \ref{prop:lb1} and $D^{(2)}(R_1,R_2,\theta_1,\theta_2)$ from Proposition \ref{prop:prop6} are also plotted for comparison. We observe that the distortion of the proposed scheme decreases as $R_2$ increases and gets arbitrarily close to lower bound $D^{(1)}(R_1,R_2,\theta_1,\theta_2)$, corresponding to the cut around Encoder 1, as $R_2$ gets sufficiently large. In contrast, lower bound $D^{(2)}(R_1,R_2,\theta_1,\theta_2)$, corresponding to the cut around the decoder, is tighter than lower bound $D^{(1)}(R_1,R_2,\theta_1,\theta_2)$ for small rates $R_2$.

The optimal overlap fraction $\theta_{12}^*$ for the achievable rate is equal to the minimum possible fraction $\theta_{12,min} = 0.25$ when $R_2$ is sufficiently small, i.e., $R_2\leq 0.03$. It is interesting to compare this result with Proposition \ref{cor:cor3}. In fact, the latter entails that, for the point-to-point case when $R_1$ goes to infinity, if rate $R_2$ is small enough, then the optimal overlap fraction $\theta_{12}^*$ equals the largest possible value $\theta_{12,max}$. Thus, the optimality of the choice $\theta_{12}^*=\theta_{12,min}$ in the multi-hop scenario with the given value of $R_1$ is due to the fact that the source $S_1$ received at Encoder 2 is noisy as a result of the necessary compression at Encoder 1. Moreover, since rate $R_2$ is very small, similar to Proposition \ref{cor:cor3}, the rates of both links are entirely allocated for describing the overlapped fraction of samples.\footnote{The fact that $\theta_{12}^*$ equals $\theta_{12,min}$, despite the fact that only the overlapped samples are described, is explained mathematically by the non-convexity of function $D_0(R_1,R_2)$.} As rate $R_2$ increases, the optimal overlap fraction $\theta_{12}^*$ gradually increases until it reaches the maximum possible value $\theta_{12,max} = 0.5$. In this regime, in addition to the overlapped fraction of samples, Encoder 2 also starts describing the non-overlapped fraction of samples measured only from source $S_2$. Finally, as $R_2$ grows beyond 0.16, the non-overlapped fraction of samples measured only from source $S_1$ also starts being allocated a non-zero rate and the optimal overlap fraction $\theta_{12}^*$ decreases down to the minimum possible value $\theta_{12,min} = 0.25$. This is consistent with Proposition \ref{prop:lb1}, which shows that as $R_2$ goes to infinity, the optimal overlap fraction is $\theta_{12,min} = 0.25$.

\section{Conclusions}

In this paper, we considered the problem of lossy compression
for computing a function of correlated sources. Motivated by the fact
that acquiring the information necessary for computation may be costly
in sensor networks, we assumed that the encoder can only observe a
fraction of the samples from each source according to a sampling strategy
that is subject to design. We also investigated the corresponding multi-hop problem with two encoders each observing a fraction of one of the sources. The results highlight the dependence of
the optimal sampling strategy on the function to be computed by the
decoder, on the source statistics, including the correlation, on the link rate and the desired metric for distortion.
Interesting future work includes investigation of other network scenarios and extensions to sources with memory.

\appendices

\section{Proof of Proposition \ref{prop:prop4}}

\label{app:gaussian_sum}Given $T = S_{1}$, we have the distortion rate functions $D_1(R) = D_{12}(R) = 2^{-2R}$ and $D_2(R) = 1-\rho^2 +\rho^2 2^{-2R}$ \cite{processingcost:Oohama_IT97}. In this case, applying Lemma \ref{lem:lem1}, we obtain
\begin{align}
&D(R,\theta_1,\theta_2,\theta_{12})\nonumber\\
 = &\min_{R_1,R_{2}, R_{12}\geq 0} (\theta_1-\theta_{12})2^{-\frac{2R_1}{\theta_1-\theta_{12}}} + \theta_{12} 2^{-\frac{2R_{12}}{\theta_{12}}}+ (\theta_2-\theta_{12})\left(1-\rho^2 + \rho^2 2^{-\frac{2R_2}{\theta_2-\theta_{12}}}\right) + (1+\theta_{12}-\theta_1-\theta_2)\label{eqn:dist_rate2}\\
= &\min_{0\leq R_2\leq R} \theta_1 2^{-\frac{2(R-R_2)}{\theta_1}} + (\theta_2-\theta_{12})\rho^2 2^{-\frac{2R_2}{\theta_2-\theta_{12}}} + 1-\theta_1 -\rho^2(\theta_2-\theta_{12}),\label{eqn:dist_rate1}
\end{align}
where the minimization in (\ref{eqn:dist_rate2}) is under the constraint $R_1+R_{2}+R_{12}\leq R$. Note that the optimization in (\ref{eqn:dist_rate2}) is equivalent to that in (\ref{eqn:dist_rate1}), since in any optimal solution, we have $R_{12}/\theta_{12} = R_1/(\theta_1-\theta_{12})$ by the convexity of function $2^{-2r}$ for $r\geq 0$, and the condition $R_1+R_{2}+R_{12}\leq R$ must be met with equality. It can be easily seen that function $D(R,\theta_1,\theta_2,\theta_{12})$ above is monotonically non-decreasing with respect to $\theta_{12}$. Therefore, the optimal overlap is the minimum possible, which equals $\theta_{12}^* = \theta_{12,min}$. Moreover, the optimal rate $R_2^*$ that minimizes (\ref{eqn:dist_rate1}) can be obtained using standard Lagrangian methods similar to \cite{processingcost:Liu_Simeone_Erkip_tcom} as:
\begin{equation}
R_{2}^* = \frac{\theta_2-\theta_{12}^*}{\theta_1+\theta_2-\theta_{12}^*}\left(R-\frac{\theta_1}{2}\log_2\frac{1}{\rho^2}\right)^+.
\end{equation}
With the so obtained $R_2^*$, the results in Proposition \ref{prop:prop4} follows immediately.

\vspace{-0.25cm}
\section{Proof of Proposition \ref{prop:prop3}}

\label{app:bin_prod_largeR}For any given sampling profile $(\theta_1,\theta_2,\theta_{12})$, in order to minimize the distortion with respect to $R$, we can take $R$ to be arbitrarily large (in fact, given the binary alphabets, $R=1$ suffices). With no rate limitations, it is easy to see that, during the $\theta_{12}$-fraction, $T$ can be computed at the encoder and described to the decoder losslessly with a rate equal to the entropy of $T$, $h((1-p)/2)$. During the $(\theta_1-\theta_{12})$-fraction, only source $S_1$ is observed and can be described to the decoder losslessly with a rate $h(1/2) = 1$. Based on source $S_1$, the best estimate at the decoder is as follows: $\hat{T} = 0$ if $S_1 = 0$, and $\hat{T} = 1$ if $S_1 = 1$, leading to average Hamming distortion $p/2$. Similarly, during the $(\theta_2-\theta_{12})$-fraction, the average Hamming distortion is also $p/2$. During the $(1+\theta_{12}-\theta_1-\theta_2)$-fraction, neither source $S_1$ nor source $S_2$ is observed. Since $T$ is Bernoulli distributed with $(1-p)/2$, the best estimate is given by $\hat{t} = 0$, leading to average Hamming distortion $(1-p)/2$. Therefore, we have $D_{1,min} = D_{2,min} = p/2$ and $D_{max} = (1-p)/2$. Substituting into (\ref{eqn:Dmin0}) of Lemma \ref{lem:lem_Dmin}, we have
\begin{align}
D_{min}(\theta_1,\theta_2)  = \frac{1-p}{2} + \left(p-\frac{1}{2}\right)(\theta_1+\theta_2) + \frac{1-3p}{2}\theta_{12}^*, \label{eqn:Dmin1}
\end{align}
where $\theta_{12}^* = \theta_{12,min}$ if $p<1/3$, and $\theta_{12}^* = \theta_{12,max}$ if $p\geq 1/3$. Finally, from the discussion above, it follows that, for any $R\geq R_{min}(\theta_1,\theta_2) = \theta_1+\theta_2 - \left(2-h((1-p)/2)\right)\theta_{12}^*$, distortion $D_{min}(\theta_{1},\theta_{2})$ can be achieved at the decoder.

\vspace{-0.25cm}
\section{Proof of Lemma \ref{lem:lem3}}

\label{app:app1}In the case of indirect description of $T$ based on only source $S_1$, the indirect rate-distortion function is given by $R_1(D) = \min_{p(\hat{t}|s_1):\;Ed(T,\hat{T})\leq D} I(S_1;\hat{T})$ \cite{processingcost:Witsenhausen_IT80}. Let $p(\hat{t} = 1|s_1 = 0) = x$ and $p(\hat{t} = 1|s_1 = 1) = y$, where $0\leq x\leq 1$ and $0\leq y\leq 1$. Note that if we select $x = y = 0$, i.e., $\hat{T} = 0$ with probability 1, the average distortion $D =(1-p)/2$ is achievable at the decoder. Thus, for $D\geq (1-p)/2$, we have $R_1(D) = 0$. Moreover, from the proof of Proposition \ref{prop:prop3}, it follows that $D\geq p/2$ must hold. For the nontrivial case $p/2\leq D< (1-p)/2$, the expected distortion constraint can be written as
\begin{align}
E(d(T,\hat{T})) = \frac{x}{2} + \frac{1-y}{2}(1-p) + \frac{y}{2}p = \frac{x+(2p-1)y+1-p}{2} \leq D, \label{eqn:avgdist}
\end{align}
and the mutual information $I(S_1;\hat{T})$ can be written as
\begin{align}
I(S_1;\hat{T}) = H(\hat{T})-H(\hat{T}|S_1)= h\left(\frac{x+y}{2}\right)-\frac{1}{2}h(x)-\frac{1}{2}h(y).\label{eqn:mutinf}
\end{align}
For any given $y$, considering the monotonicity of (\ref{eqn:mutinf}) with respect to $x$ for $0\leq x\leq 2D-(1-p)+(1-2p)y$, we can easily show that (\ref{eqn:mutinf}) is minimized at $x = 2D-(1-p)+(1-2p)y$, i.e., (\ref{eqn:avgdist}) is met with equality. Therefore, for $p/2\leq D< (1-p)/2$, we obtain $R_1(D)$ as
in (\ref{eqn:R1_D}).

\vspace{-0.25cm}
\section{Proof of Proposition \ref{prop:prop5}}

\label{app:bin_prod}If we set $\hat{T} = 0$ at the decoder, the resulting Hamming distortion is $(1-p)/2$. Hence, for $D\geq (1-p)/2$, zero rate is required for description, i.e., $R(D,\theta_1,\theta_2) = 0$. For $D_{min}(\theta_{1},\theta_{2})\leq D<\frac{1-p}{2}$, for any given sampling profile $(\theta_1,\theta_2,\theta_{12})$, we can use Lemma \ref{lem:lem1} by setting $D_{1,min} = D_{2,min} = p/2$, $D_{12,min} = 0$ and $D_{max} = (1-p)/2$. Due to the convexity of $R_1(D)$, it is optimal to have $D_1/(\theta_1-\theta_{12}) =D_2 /(\theta_2-\theta_{12})$ in any optimal solution. Moreover, with $D_{min}(\theta_{1},\theta_{2})\leq D<\frac{1-p}{2}$, for optimality, (\ref{eqn:dist_sum}) must be met with equality, i.e.,
\begin{equation}
D_1+D_2+D_{12}+\frac{(1+\theta_{12}-\theta_1-\theta_2)(1-p)}{2}=D,
\end{equation}
and $D_{1}$, $D_{12}$ and $D_{2}$ must be such that $D_1/(\theta_1-\theta_{12})$, $D_{12}/\theta_{12}$ and $D_2/(\theta_2-\theta_{12})$ are all less than or equal to $D_{max} = (1-p)/2$. If we let $D_3 = D_1+ D_2$, then $D_3$ satisfies
\begin{equation}
\frac{p(\theta_1+\theta_2-2\theta_{12})}{2}\leq D_3\leq \frac{(1-p)(\theta_1+\theta_2-2\theta_{12})}{2}.
\end{equation}
Finally, taking the minimum of $R(D,\theta_1,\theta_2,\theta_{12})$ over all $\theta_{12}$ satisfying (\ref{eqn:cnst_theta12}), we obtain $R(D,\theta_1,\theta_2)$ as in the proposition for $D_{min}(\theta_{1},\theta_{2})\leq D<\frac{1-p}{2}$.
\vspace{-0.25cm}

\section{Proof of Proposition \ref{prop:lb1}}
\label{app:lowerbound1}
For any given sampling profile $(\theta_1,\theta_2,\theta_{12})$, we denote by $R_{11}$ and $R_{12}$, the rates used by the encoder in Fig. \ref{fig:p2p1} to describe the non-overlapping $(\theta_1-\theta_{12})$-fraction of samples and the overlapping $\theta_{12}$-fraction of samples measured from $S_1$, respectively. During the $\theta_{12}$-fraction, using rate $R_{12}/\theta_{12}$, one can achieve average distortion $(1-\rho^2)2^{-2R_{12}/\theta_{12}}$ by the Wyner-Ziv theorem \cite{processingcost:Wyner78}. During the $(\theta_1-\theta_{12})$-fraction, only source $S_1$ is observed and described to the decoder using rate $R_{11}/(\theta_1-\theta_{12})$ and the resulting average distortion is given by $2(1+\rho)(1-\tilde{\rho}^2+\tilde{\rho}^2 2^{-2R_{11}/(\theta_1-\theta_{12})})$, where $\tilde{\rho}$ is as defined in Section IV. During the $(\theta_2-\theta_{12})$-fraction, since only $S_2$ is observed perfectly at the decoder, the resulting average distortion can be easily seen to be $1-\rho^2$. Finally, during the $(1+\theta_{12}-\theta_1-\theta_2)$-fraction, with neither source $S_1$ nor source $S_2$ observed, the average distortion at the decoder is equal to the variance of $T$, namely, $2(1+\rho)$. From the independence of samples measured from the different fractions of samples, the minimum achievable distortion for sampling budget $(\theta_1,\theta_2)$ and rate $R_1$ can be obtained as
\vspace{-0.15cm}\begin{align}
D^{(1)}(R_1,\theta_1,\theta_2)= & \min_{\theta_{12},R_{11},R_{12}}(1-\rho^2)\theta_{12} 2^{-\frac{2R_{12}}{\theta_{12}}} + (1+\rho)^2(\theta_1-\theta_{12}) 2^{-\frac{2R_{11}}{\theta_1-\theta_{12}}} \nonumber\\
 & +2\rho(1+\rho)\theta_{12}-(1+\rho)^2(\theta_1+\theta_2)+2(1+\rho) , \label{eqn:dis_LB2}
\end{align}
\vspace{-0.15cm}where the constraint on $\theta_{12}$ is as in (1) and the constraint on $R_{11},R_{12}$ is given by $R_{11}+R_{12}\leq R_1$. The minimum achievable distortion in the proposition is obtained by solving the optimization problem (\ref{eqn:dis_LB2}). Specifically, for $\rho>0$, we can show that it is optimal to have $\theta_{12}^* = \theta_{12,min}$ by simply considering the monotonicity of function $D^{(1)}(R,\theta_1,\theta_2,\theta_{12})$ with respect to $\theta_{12}$. Similarly, we can show that for $\rho\leq 0$, it is optimal to have $\theta_{12}^* = \theta_{12,max}$. Moreover, the optimal rate $R_{11}^*$ and $R_{12}^*$ can be obtained using standard Lagrangian methods similar to \cite{processingcost:Liu_Simeone_Erkip_tcom}. Details are omitted.

\section{Trade-Off between Average Distortion and Worst-Case Distortion}
\label{app:worst}
In the problem formulation considered in Section II, the goal was minimizing the average distortion $D(R,\theta_1,\theta_2)$ at the decoder for any given rate $R$ and sampling budget $(\theta_{1},\theta_{2})$. As a result of the average of over all the source samples in (\ref{eqn:AvgDist0}), this performance metric does not make guarantees on the maximum average distortion per sample. In fact, as seen in (\ref{eqn:D_lemma1}), the average distortion is the average of the distortions accrued over the four relevant fractions of samples illustrated in Fig. \ref{fig:fraction}, namely the fraction of samples in which only $S_1$, both $S_1$ and $S_2$, only $S_2$ or neither $S_1$ nor $S_2$ are measured.  In this appendix, we extend the analysis in Sections III, IV and V in order to allow the decoder to strike the desired balance between the average and the worst-case distortions.

\vspace{-0.25cm}\subsection{Formulation for General Sources}
For any given sampling profile $(\theta_{1},\theta_{2},\theta_{12})$ and any rate allocation $(R_1,R_2,R_{12})$, we define $D_{w}(R_1,R_2,R_{12},\theta_{1},\theta_{2},\theta_{12})$ as the maximum average distortion among all the sampling fractions shown in Fig. \ref{fig:fraction}, i.e.,
\begin{align}
D_{w}(R_1,R_2,R_{12},\theta_{1},\theta_{2},\theta_{12}) = &\max\left[1_{\{\theta_1-\theta_{12}>0\}}D_1\left(\frac{R_1}{\theta_1-\theta_{12}}\right), 1_{\{\theta_{12}>0\}}D_{12}\left(\frac{R_{12}}{\theta_{12}}\right),\right.\nonumber\\
&\left. 1_{\{\theta_2-\theta_{12}>0\}}D_2\left(\frac{R_2}{\theta_2-\theta_{12}}\right), 1_{\{1+\theta_{12}-\theta_1-\theta_2>0\}}D_{max}\right],\label{eqn:worstCaseDist}
\end{align}
where the indicator function $1_{\{A\}}$ takes value 1 if $A$ is true and 0 otherwise. We then define the weighted sum of the average distortion in (\ref{eqn:AvgDist0}) and the worst-case distortion (\ref{eqn:worstCaseDist}) as
\begin{align}
D_{\mu}(R,\theta_1,\theta_2,\theta_{12}) = & \min_{R_1,R_{2},R_{12}\geq 0}(\theta_1-\theta_{12})D_1\left(\frac{R_1}{\theta_1-\theta_{12}}\right) + \theta_{12}D_{12}\left(\frac{R_{12}}{\theta_{12}}\right)+(\theta_2-\theta_{12})D_2\left(\frac{R_2}{\theta_2-\theta_{12}}\right)\nonumber\\
& + (1+\theta_{12}-\theta_1-\theta_2)D_{max} + \mu D_{w}(R_1,R_2,R_{12},\theta_{1},\theta_{2},\theta_{12}),\label{eqn:modified_dist}
\end{align}
where $\mu>0$ is the relative weight of the worst-case distortion.
Accordingly, given any sampling budget $(\theta_{1},\theta_{2})$, the modified distortion-rate trade-off is characterized by $D_{\mu}(R,\theta_1,\theta_2)= \inf_{\theta_{12}}D_{\mu}(R,\theta_1,\theta_2,\theta_{12})$.

In the case of sampling budget $(\theta_1,\theta_2)$ satisfying $\theta_{1} + \theta_{2}<1$, regardless of the choice of the overlap fraction $\theta_{12}$ and of the rate allocation $(R_1,R_2,R_{12})$, the worst-case distortion occurs during the $(1+\theta_{12}-\theta_1-\theta_2)$-fraction of samples in which neither source is measured and therefore we have $D_{w}(R_1,R_2,R_{12},\theta_{1},\theta_{2},\theta_{12}) = D_{max}$. In this case, the addition of a constant term in the objective function of (\ref{eqn:modified_dist}) does not affect the optimization and thus the optimal overlap fraction $\theta_{12}^*$ and the optimal rate allocation $(R_1^*,R_{2}^*,R_{12}^*)$ remains the same as in the case when $\mu = 0$. Therefore, in the following, we focus on the nontrivial case where we have $0<\theta_1<1$, $0<\theta_2<1$ and $\theta_1+\theta_2\geq 1$. In this regime, function $D_{\mu}(R,\theta_1,\theta_2,\theta_{12})$ takes different forms depending on the value of $\theta_{12}$. Specifically, if the overlap fraction $\theta_{12}$ is selected such that $\theta_{12,min}<\theta_{12}\leq \theta_{12,max}$, then we have
\begin{align}
D_{\mu}(R,\theta_1,\theta_2,\theta_{12}) = & \min_{R_1,R_{12},R_2\geq 0}(\theta_1-\theta_{12})D_1\left(\frac{R_1}{\theta_1-\theta_{12}}\right) + \theta_{12}D_{12}\left(\frac{R_{12}}{\theta_{12}}\right)\nonumber\\
&+(\theta_2-\theta_{12})D_2\left(\frac{R_2}{\theta_2-\theta_{12}}\right) + (1+\theta_{12}-\theta_1-\theta_2)D_{max} + \mu D_{max}.\label{eqn:modified_dist2}
\end{align}
This is because, with $\theta_{12,min} = \theta_1+\theta_2-1$, we have $1+\theta_{12}-\theta_1-\theta_2>0$ and accordingly $D_{w}(\theta_{1},\theta_{2},\theta_{12}, R_1,R_2,R_{12}) = D_{max}$. The minimum distortion (\ref{eqn:modified_dist2}) when $\theta_{12}$ is in the interval $\theta_{12,min}<\theta_{12}\leq \theta_{12,max}$ thus can be obtained in the same manner as done in the previous sections when $\mu = 0$. On the other hand, if the overlap fraction is $\theta_{12} = \theta_{12,min} = \theta_1+\theta_2-1$, then we can write
\begin{align}
D_{\mu}&(R,\theta_1,\theta_2,\theta_{12,min}) =  \min_{R_1,R_{12},R_2}(1-\theta_2)D_1\left(\frac{R_1}{1-\theta_{2}}\right) + (\theta_{1}+\theta_{2}-1)D_{12}\left(\frac{R_{12}}{\theta_{1}+\theta_{2}-1}\right)\nonumber\\
& +(1-\theta_{1})D_2\left(\frac{R_2}{1-\theta_{1}}\right) + \mu \max\left(D_1\left(\frac{R_1}{1-\theta_2}\right),1_{\{\theta_1+\theta_2-1>0\}}D_{12}\left(\frac{R_{12}}{\theta_1+\theta_2-1}\right), D_2\left(\frac{R_2}{1-\theta_1}\right)\right).\label{eqn:tmp2}
\end{align}
The minimization (\ref{eqn:tmp2}) is discussed below for the Gaussian case. The minimization of (\ref{eqn:modified_dist}) is then obtained by taking the minimum between (\ref{eqn:tmp2}) and the value obtained for the interval $\theta_{12,min}<\theta_{12}\leq \theta_{12,max}$  from (\ref{eqn:modified_dist2}).
\begin{figure}
\centering
\includegraphics[width = 3in]{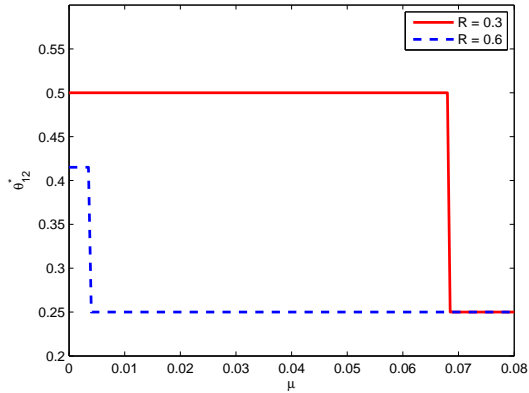}
\caption{Optimal overlap fraction $\theta_{12}^*$ that minimizes distortion $D_{\mu}(R,\theta_1,\theta_2)$ as a function of $\mu$ for computing $T=S_1+S_2$, $(S_1,S_2)$ jointly Gaussian with correlation coefficient $\rho = 0.5$. The sample budget is $(\theta_1,\theta_2) = (0.5,0.75)$ and the link rate is $R = 0.3, 0.6$, respectively.}
\label{fig:optTheta12VSmu}
\end{figure}

\subsection {Computation of the Sum of Jointly Gaussian Sources}
In the special case of Gaussian sources $S_1$ and $S_2$ and  of calculation of the function $T = S_1 + S_2$ as in Section IV-B, we can simplify (\ref{eqn:tmp2}) as
\begin{align}
D_{\mu}(R,\theta_1,\theta_2,\theta_{12,min})
= & \min_{R_1,R_{12},R_2}(2-\theta_1-\theta_2)D_1\left(\frac{R_1+R_2}{2-\theta_1-\theta_{2}}\right) + (\theta_{1}+\theta_{2}-1)D_{12}\left(\frac{R_{12}}{\theta_{1}+\theta_{2}-1}\right)\nonumber\\
&+ \mu \max\left(D_1\left(\frac{R_1+R_2}{2-\theta_1-\theta_2}\right),1_{\{\theta_1+\theta_2-1>0\}}D_{12}\left(\frac{R_{12}}{\theta_1+\theta_2-1}\right)\right)\label{eqn:tmp1}\\
= &\min_{R_{12}}(2-\theta_1-\theta_2)D_1\left(\frac{R-R_{12}}{2-\theta_1-\theta_{2}}\right) + (\theta_{1}+\theta_{2}-1)D_{12}\left(\frac{R_{12}}{\theta_{1}+\theta_{2}-1}\right)\nonumber\\
&+ \mu \max\left(D_1\left(\frac{R-R_{12}}{2-\theta_1-\theta_2}\right),1_{\{\theta_1+\theta_2-1>0\}} D_{12}\left(\frac{R_{12}}{\theta_1+\theta_2-1}\right)\right),\label{eqn:tmp3}
\end{align}
where (\ref{eqn:tmp1}) follows from the facts that $D_1(R) = D_2(R)$ for $R\geq 0$ and that the function is minimized when $R_1/(1-\theta_2)=R_2/(1-\theta_1)$; and (\ref{eqn:tmp3}) follows because the constraint $R_1+R_2+R_{12}\leq R$ is easily seen to be met with equality in any optimal solution.

We now numerically evaluate (\ref{eqn:tmp3}), the trade-off between the average distortion and the worst distortion for the Gaussian case at hand. Fig. \ref{fig:optTheta12VSmu} shows the optimal overlap fraction $\theta_{12}^*$ versus the relative weight $\mu$ when $(\theta_1, \theta_2) = (0.5, 0.75)$ and $\rho = 0.5$, for two choices of rates $R = 0.3$ and $R = 0.6$ respectively. From the curve with $R = 0.3$, we can see that, for sufficiently small $\mu$, here, $\mu\leq 0.069$, the optimal sampling fraction is $\theta_{12,max} = 0.5$, consistent with the result of Fig. \ref{fig:opt_theta12}. However, as $\mu$ grows larger, the optimal sampling fraction drops to the minimum possible value $\theta_{12,min} = 0.25$. This is since, as the value of $\mu$ grows sufficiently large, one needs to keep the worst distortion as small as possible. Similar conclusions are reached for $R = 0.6$, except that the optimal sampling fraction for sufficiently small $\mu$ ($\mu\leq 0.004$) is $0.42$ instead, which is also consistent with the result of Fig. \ref{fig:opt_theta12}.

\bibliographystyle{IEEEtran}
\bibliography{IEEEabrv,processingcost}

\end{document}